\documentclass[12pt,dvips,oneside]{revtex4}
\usepackage{amsmath,amssymb,graphicx}

\newcommand{\bfx} { {\bf X} }

\newcommand{\bft} { \hat{\mbox{\boldmath$\tau$}} }
\newcommand{\bflp}{ \mbox{\boldmath$\bigtriangledown$} }

\newcommand{\gt}{ {\bf g}  }
\newcommand{\gtpar}{ {\bf g}^{\parallel} }
\newcommand{\gtper}{ {\bf g}^{\perp}     }
\newcommand{\hgtper}{ {\hat{\bf g}}^{\perp}     }

\newcommand{\gs}{ \widetilde{{\bf g}}     }
\newcommand{\gspar}{ \widetilde{ {\bf g}}^{\parallel} }
\newcommand{\gsper}{ \widetilde{ {\bf g}}^{\perp}     }

\newcommand{\hess}{{\bf H}}

\begin{document}
\def\aciee{Angew. Chem. Int. Ed. Engl. }
\def\ac{Acta. Crystallogr. }
\def\acp{Adv. Chem. Phys. }
\def\acr{Acc. Chem. Res. }
\def\ajp{Am. J. Phys. }
\def\ap{Ann. Physik }
\def\apc{Adv. Prot. Chem. }
\def\arpc{Ann. Rev. Phys. Chem. }
\def\cccc{Coll. Czech. Chem. Comm. }
\def\cpc{Comp. Phys. Comm. }
\def\cpl{Chem. Phys. Lett. }
\def\crev{Chem. Rev. }
\def\el{Europhys. Lett. }
\def\ic{Inorg. Chem. }
\def\ijmpc{Int. J. Mod. Phys. C }
\def\ijqc{Int. J. Quant. Chem. }
\def\jcis{J. Colloid Interface Sci. }
\def\jcsft{J. Chem. Soc., Faraday Trans. }
\def\jacs{J. Am. Chem. Soc. }
\def\jas{J. Atmos. Sci. }
\def\jbc{J. Biol. Chem. }
\def\jcc{J. Comp. Chem. }
\def\jcp{J. Chem. Phys. }
\def\jce{J. Chem. Ed. }
\def\jcscc{J. Chem. Soc., Chem. Commun. }
\def\jetp{J. Exp. Theor. Phys. (Russia) }
\def\jmb{J. Mol. Biol. }
\def\jmsp{J. Mol. Spec. }
\def\jmst{J. Mol. Struct. }
\def\jncs{J. Non-Cryst. Solids }
\def\jpa{J. Phys. A }
\def\jpc{J. Phys. Chem. }
\def\jpca{J. Phys. Chem. A }
\def\jpcb{J. Phys. Chem. B }
\def\jpcm{J. Phys. Condensed Matter. }
\def\jpcs{J. Phys. Chem. Solids. }
\def\jpsj{J. Phys. Soc. Jpn. }
\def\jrnist{J. Res. Natl. Inst. Stand. Technol. }
\def\mg{Math. Gazette }
\def\mp{Mol. Phys. }
\def\nat{Nature }
\def\nsb{Nat. Struct. Biol.}
\def\Pa{Physica A }
\def\pac{Pure. Appl. Chem. }
\def\pccp{Phys. Chem. Chem. Phys. }
\def\phys{Physics }
\def\pmb{Philos. Mag. B }
\def\ptrsb{Philos. T. Roy. Soc. B }
\def\pnasu{Proc. Natl. Acad. Sci. USA }
\def\pr{Phys. Rev. }
\def\prep{Phys. Reports }
\def\pra{Phys. Rev. A }
\def\prb{Phys. Rev. B }
\def\prbcm{Phys. Rev. B }
\def\prc{Phys. Rev. C }
\def\prd{Phys. Rev. D }
\def\pre{Phys. Rev. E }
\def\prl{Phys. Rev. Lett. }
\def\prsa{Proc. R. Soc. A }
\def\psfg{Proteins: Struct., Func. and Gen. }
\def\sci{Science }
\def\spj{Sov. Phys. JETP }
\def\ss{Surf. Sci. }
\def\tca{Theor. Chim. Acta }
\def\zpb{Z. Phys. B. }
\def\zpc{Z. Phys. Chem. }
\def\zpd{Z. Phys. D }
\def\aciee{Angew. Chem. Int. Ed. Engl. }
\def\ac{Acta. Crystallogr. }
\def\acp{Adv. Chem. Phys. }
\def\acr{Acc. Chem. Res. }
\def\ajp{Am. J. Phys. }
\def\ap{Adv. Phys. }
\def\arpc{Ann. Rev. Phys. Chem. }
\def\cccc{Coll. Czech. Chem. Comm. }
\def\cpl{Chem. Phys. Lett. }
\def\crev{Chem. Rev. }
\def\dalton{J. Chem. Soc., Dalton Trans. }
\def\el{Europhys. Lett. }
\def\faraday{J. Chem. Soc., Faraday Trans. }
\def\fartrans{J. Chem. Soc., Faraday Trans. }
\def\fdisc{J. Chem. Soc., Faraday Discuss. }
\def\ic{Inorg. Chem. }
\def\ijqc{Int. J. Quant. Chem. }
\def\jcis{J. Colloid Interface Sci. }
\def\jcsft{J. Chem. Soc., Faraday Trans. }
\def\jacs{J. Am. Chem. Soc. }
\def\jas{J. Atmos. Sci. }
\def\jcc{J. Comp. Chem. }
\def\jcp{J. Chem. Phys. }
\def\jce{J. Chem. Ed. }
\def\jcscc{J. Chem. Soc., Chem. Commun. }
\def\jetp{J. Exp. Theor. Phys. (Russia) }
\def\jmc{J. Math. Chem. }
\def\jmsp{J. Mol. Spec. }
\def\jmst{J. Mol. Structure }
\def\jncs{J. Non-Cryst. Solids }
\def\jpc{J. Phys. Chem. }
\def\jpcm{J. Phys. Condensed Matter. }
\def\jpsj{J. Phys. Soc. Jpn. }
\def\jsp{J. Stat. Phys. }
\def\mg{Math. Gazette }
\def\mp{Mol. Phys. }
\def\molphys{Mol. Phys. }
\def\nat{Nature }
\def\pac{Pure. Appl. Chem. }
\def\phys{Physics }
\def\pla{Phys. Lett. A }
\def\plb{Phys. Lett. B }
\def\phm{Philos. Mag. }
\def\pmb{Philos. Mag. B }
\def\pnas{Proc.\ Natl.\ Acad.\ Sci.\  USA }
\def\pr{Phys. Rev. }
\def\pra{Phys. Rev. A }
\def\prb{Phys. Rev. B }
\def\prc{Phys. Rev. C }
\def\prd{Phys. Rev. D }
\def\pre{Phys. Rev. E }
\def\prl{Phys. Rev. Lett. }
\def\prsa{Proc. R. Soc. A }
\def\ss{Surf. Sci. }
\def\sci{Science }
\def\tca{Theor. Chim. Acta }
\def\zpc{Z. Phys. Chem. }
\def\zpd{Z. Phys. D }
\def\zfpd{Z. Phys. D }
\def\zpdamc{Z. Phys. D }
\def\aciee{Angew. Chem. Int. Ed. Engl. }
\def\ac{Acta. Crystallogr. }
\def\acp{Adv. Chem. Phys. }
\def\acr{Acc. Chem. Res. }
\def\ajp{Am. J. Phys. }
\def\am{Adv. Mater. }
\def\apl{Appl. Phys. Lett. }
\def\arpc{Ann. Rev. Phys. Chem. }
\def\mrsb{Mater. Res. Soc. Bull. }
\def\cccc{Coll. Czech. Chem. Comm. }
\def\cj{Comput. J. }
\def\cp{Chem. Phys. }
\def\cpc{Comp. Phys. Comm. }
\def\cpl{Chem. Phys. Lett. }
\def\crev{Chem. Rev. }
\def\el{Europhys. Lett. }
\def\fd{Faraday Disc. }
\def\ic{Inorg. Chem. }
\def\ijmpc{Int. J. Mod. Phys. C }
\def\ijqc{Int. J. Quant. Chem. }
\def\jcis{J. Colloid Interface Sci. }
\def\jcsft{J. Chem. Soc., Faraday Trans. }
\def\jacs{J. Am. Chem. Soc. }
\def\jap{J. Appl. Phys. }
\def\jas{J. Atmos. Sci. }
\def\jcc{J. Comp. Chem. }
\def\jcp{J. Chem. Phys. }
\def\jce{J. Chem. Ed. }
\def\jcscc{J. Chem. Soc., Chem. Commun. }
\def\jetp{J. Exp. Theor. Phys. (Russia) }
\def\jmsp{J. Mol. Spec. }
\def\jmst{J. Mol. Structure }
\def\jncs{J. Non-Cryst. Solids }
\def\jpa{J. Phys. A }
\def\jpc{J. Phys. Chem. }
\def\jpcssp{J. Phys. C: Solid State Phys. }
\def\jpca{J. Phys. Chem. A. }
\def\jpcb{J. Phys. Chem. B. }
\def\jpcm{J. Phys. Condensed Matter. }
\def\jpcs{J. Phys. Chem. Solids. }
\def\jpsj{J. Phys. Soc. Jpn. }
\def\jpfmp{J. Phys. F, Metal Phys. }
\def\mg{Math. Gazette }
\def\mp{Mol. Phys. }
\def\msr{Mater. Sci. Rep. }
\def\nat{Nature }
\def\njc{New J. Chem. }
\def\pac{Pure. Appl. Chem. }
\def\phys{Physics }
\def\pma{Philos. Mag. A }
\def\pmb{Philos. Mag. B }
\def\pml{Philos. Mag. Lett. }
\def\pnasu{Proc. Natl. Acad. Sci. USA }
\def\pr{Phys. Rev. }
\def\prep{Phys. Reports }
\def\pra{Phys. Rev. A }
\def\prb{Phys. Rev. B }
\def\prc{Phys. Rev. C }
\def\prd{Phys. Rev. D }
\def\pre{Phys. Rev. E }
\def\prl{Phys. Rev. Lett. }
\def\prsa{Proc. R. Soc. A }
\def\pss{Phys. State Solidi }
\def\pssb{Phys. State Solidi B }
\def\rmp{Rev. Mod. Phys. }
\def\rpp{Rep. Prog. Phys. }
\def\sci{Science }
\def\ss{Surf. Sci. }
\def\tca{Theor. Chim. Acta }
\def\tetra{Tetrahedron }
\def\zpb{Z. Phys. B. }
\def\zpc{Z. Phys. Chem. }
\def\zpd{Z. Phys. D }

\title{A Doubly Nudged Elastic Band Method for Finding Transition States}
\author{Semen A. Trygubenko\footnote{E-mail: sat39@cam.ac.uk} and David J. Wales}
\affiliation{University Chemical Laboratories, Lensfield Road, Cambridge CB2 1EW, UK}

\begin{abstract}
\vskip 1cm
A modification of the nudged elastic band (NEB) method is presented that 
enables stable optimisations to be run using both the 
limited-memory quasi-Newton (L-BFGS) and slow-response quenched velocity Verlet (SQVV) minimisers.
The performance of this new `doubly nudged' DNEB method 
is analysed in conjunction with both minimisers and compared with previous NEB formulations.
We find that the fastest DNEB approach
(DNEB/L-BFGS) can be quicker by up to two orders of magnitude.
Applications to permutational rearrangements of the seven-atom Lennard-Jones 
cluster (LJ$_7$) and highly cooperative
rearrangements of LJ$_{38}$ and LJ$_{75}$ are presented. We also outline an 
updated algorithm for constructing complicated multi-step pathways using successive DNEB runs.
\end{abstract}

\maketitle
\newpage

\section{Introduction}

Locating transition states on a potential energy surface (PES) provides an important tool 
in the study of dynamics using statistical rate theories \cite{Eyring35,EvansP35,Laidler87}.
Here we define a transition state according to the geometrical definition of Murrell and
Laidler, i.e.~as a stationary point with a single negative Hessian eigenvalue \cite{murrelll68}.
Unfortunately, it is significantly harder to locate transition states than local minima, since
the system must effectively `balance on a knife-edge' in one degree of freedom.
Many algorithms have been suggested for this purpose, and the most efficient method may depend
upon the nature of the system. For example, different considerations probably apply if 
second derivatives can be calculated relatively quickly, as for many empirical potentials 
\cite{wales03}.
Transformation to an alternative coordinate system may also be beneficial for systems bound by
strongly directional forces 
\cite{PulayFPB79,FogorasiZTP92,PulayF96,BakerKD96,pengasf96,bakerkp99,BilleterTT00,PaizsBSP00,BakkenH02}. 

Algorithms to locate transition states can generally be divided into single-ended approaches,
which simply require a single starting geometry
\cite{crippens71,mciverk72,pancir74,hilderbrandt77,cerjanm81,
simonsjto83,onealts84,BellC84,banerjeeass85,baker86,
NguyenC85,baker87,smith90,nicholstss90,bakerh91,baker92,tsaij93b,wales93d,chekmarev94,
wales94a,walesu94,
sunr94a,sunr94b,jensen95,bofillc95,zhaon96,quapp96,bakerc96,walshw96,ayalas97,
munrow97,ulitskys97,paizsfp98,farkass98,mousseaub98,goto98,quapphih98,bakerkp99,FarkasS99,
munrow99,walesdmmw00,kumedamw01},
and double-ended methods, which are usually designed to find a transition state between
two endpoints
\cite{Pratt86,elberk87,berrydb88,czerminskie90,FischerK92,StachoB92,
ionovac93,StachoB93,DomotorBS93,PengS93,matrofd94,smart94,BanDS94,MillsJ94,
ionovac95a,pengasf96,jonssonmj98,StachoDB99,ElberK99,
henkelmanj99,henkelmanuj00,henkelmanj00,HenkelmanJ01,MironF01,maragakisabrk02}.
The result of a single-ended search may be a transition state that is not connected to the
starting point by a steepest-descent path, and such methods can be useful for building up
databases of stationary points to provide a non-local picture of the potential energy surface,
including thermodynamic and dynamic properties \cite{walesdmmw00,wales03}.
In contrast, double-ended methods are usually designed
to characterise a particular rearrangement, and often do not produce a tightly converged geometry
for the transition state.
However, the resulting structures can be further refined using single-ended strategies,
particularly eigenvector-following 
\cite{crippens71,pancir74,hilderbrandt77,cerjanm81,simonsjto83,banerjeeass85,baker86,wales94a,
walesu94,munrow97,munrow99,kumedamw01,walesw96},
and this approach has been used in several previous studies \cite{henkelmanuj00,wales02,evansw03}.
Some recent overviews of the field are available \cite{Schlegel03,wales03}, 
and readers are referred to these publications for further discussion.




Our principal concern in the present work is the development of the double-ended NEB 
approach \cite{jonssonmj98,henkelmanj00,henkelmanuj00,henkelmanjj00}.
The earliest double-ended methods were probably the linear and quadratic synchronous
transit algorithms (LST and QST) \cite{halgrenl77}, which are entirely based on interpolation
between the two endpoints. 
In LST the highest energy structure is located along the straight line that links the two endpoints.
QST is similar in spirit, but approximates the reaction path using a parabola instead of a straight line. 
Neither interpolation is likely to provide a good estimate of the path except for 
very simple reactions, but they may nevertheless be useful to
generate initial guesses for more sophisticated double-ended methods.

Another approach 
is to reduce the distance between reactant and product by some
arbitrary value to generate an `intermediate', and seek the minimum energy
of this intermediate structure subject to certain constraints, such as fixed distance to an endpoint.
This is the basis of the `Saddle' optimisation method~\cite{dewarhs84} 
and the `Line Then Plane'~\cite{cardenas-lailhacarz95} algorithm, which differ only in the
definition of the subspace in which the intermediate is allowed to move. 
The latter method optimises the intermediate in the hyperplane
perpendicular to the interpolation line, while `Saddle' uses hyperspheres.
The minimised intermediate then replaces one of the endpoints and the process is repeated. 

There are also a number of methods that are based on a `chain-of-states' (CS) approach, where
several images of the system are somehow coupled together to create
an approximation to the required path. The CS methods mainly differ in the way in which the initial guess
to the path is refined. In the `Chain' method~\cite{liotard92} the geometry of the highest energy
image is relaxed first using only the component of the gradient perpendicular to the line
connecting its two neighbours. The process is then repeated for the next-highest
energy neighbours. The optimisation is terminated when the gradient becomes tangential
to the path. The `Locally Updated Planes' method~\cite{choie91}
is similar, but the images are relaxed
in the hyperplane perpendicular to the reaction coordinate, rather than along the line defined by the
gradient, and all the images are moved simultaneously.

The nudged elastic band (NEB) approach introduced some further refinements to these
CS methods~\cite{henkelmanjj00}.
It is based on a discretised representation of the path
originally proposed by Elber and Karplus~\cite{elberk87}, with
modifications to eliminate corner-cutting and sliding-down problems~\cite{jonssonmj98},
and to improve the stability and convergence properties~\cite{henkelmanuj00}.
Maragakis {\em et al.} applied the NEB method to various
physical systems ranging from semiconductor materials to biologically relevant molecules. They report
that use of powerful minimisation methods in conjunction with NEB approach was 
unsuccessful~\cite{maragakisabrk02}.
These problems were attributed to instabilities with
respect to the extra parameters introduced by the springs. 
In fact, the NEB approach has previously been used with the 
L-BFGS algorithm in other work, where hybrid
eigenvector-following techniques were employed to produce tightly converged 
transition states from guesses obtained by NEB calculations~\cite{wales02,evansw03}.
The main result of the present contribution is a modified `doubly nudged' elastic band (DNEB)
method, which is stable when combined with the L-BFGS minimiser.
In comparing the DNEB approach with other methods we have also analysed quenched velocity Verlet
minimisation, and determined the best point at which to remove the kinetic energy.
Extensive tests show that the DNEB/L-BFGS combination provides a significant performance improvement
over previous implementations. We therefore outline a new strategy to connect distant minima, which is
based on successive DNEB searches to provide transition state candidates for refinement
by eigenvector-following. 

\section{Methods}
\label{methods}
\begin{figure}
\centerline{\includegraphics{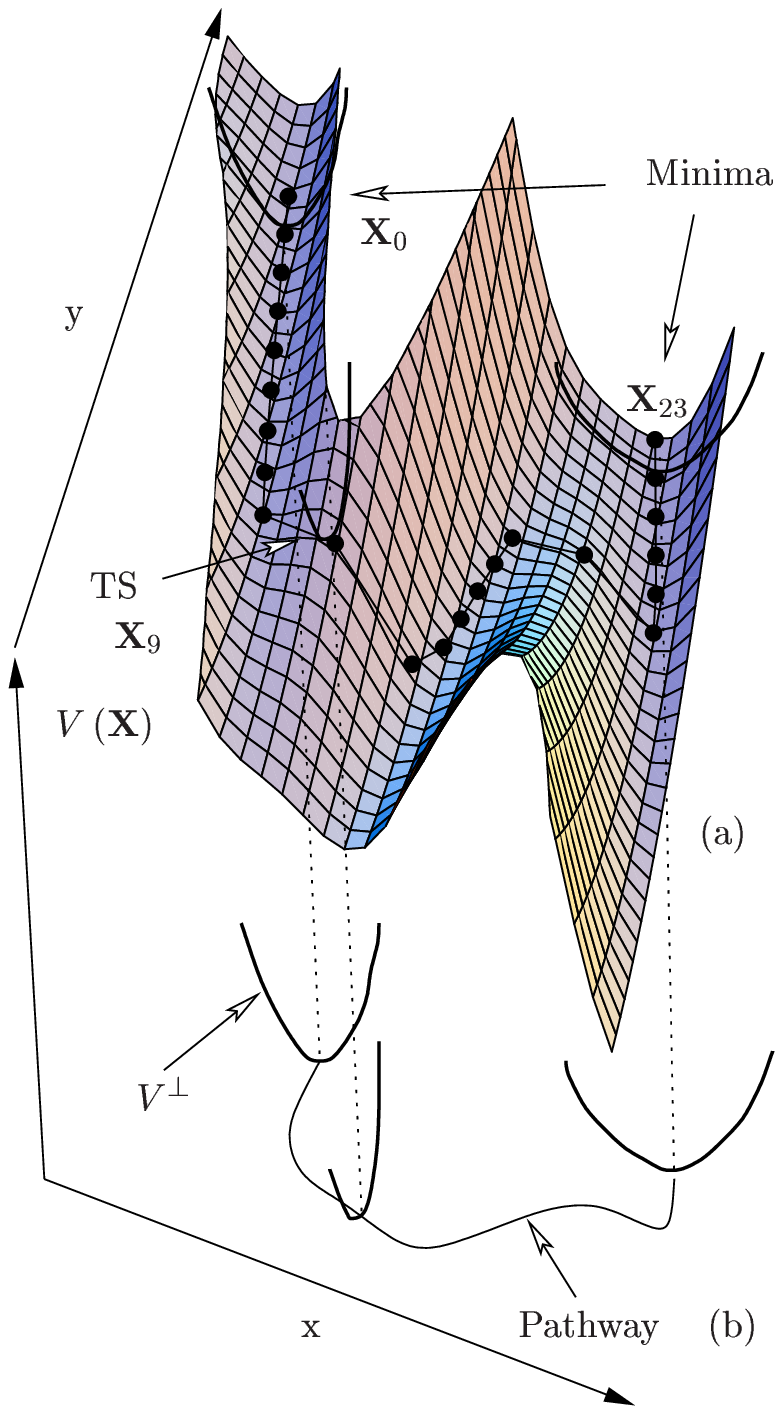}}
\caption{}
\label{fig:surface}
\end{figure}

In the present work we used the nudged elastic band~\cite{henkelmanj00,jonssonmj98} (NEB)
and eigenvector-following~\cite{crippens71,pancir74,hilderbrandt77,cerjanm81,simonsjto83,banerjeeass85,baker86,
wales94a,walesw96,munrow99,kumedamw01}
(EF) methods for locating and refining transition states.
In the NEB approach the path is represented as a set of images $\left\{\bfx_1,\bfx_2 ... \bfx_N\right\}$
that connect the endpoints $\bfx_0$ and $\bfx_{N+1}$,
where $\bfx_i$ is a vector containing the coordinates of image $i$ (Figure \ref{fig:surface})
\cite{henkelmanuj00}.
In the usual framework of double-ended methodologies~\cite{jensen99} the endpoints are stationary points
on the potential energy surface (PES) (usually minima), which are known in advance.
In addition to the true potential, $V_i$, which binds the atoms within
each image, equivalent atoms in $N$ adjacent images are interconnected by 
$N+1$ springs according to a parabolic
potential,
\begin{equation}
 	\widetilde{V} = \tfrac{1}{2} k_{spr} \sum_{i=1}^{N+1} \vert\bfx_i - \bfx_{i-1} \vert^2.
\end{equation}
Subsequently these potentials will be referred to as the 
`true potential' and the `spring potential', respectively.

The springs are intended to hold images on the path during optimisation---otherwise
they would slide down to the endpoints, or to other intermediate minima~\cite{elberk87}.
Occasionally, depending on the quality of the initial guess,
we have found that
some images may converge to higher index stationary points. One could
imagine the whole construction as a band or rope that is stretched across the
PES, which, if optimised, is capable of closely following a curve defined in terms of successive
minima, transition states, and the intervening steepest-descent paths.

In practice, the above formulation encounters difficulties connected
with the coupling between the `true' and `spring' components of the
potential.
The magnitude of the springs' interference with the true potential is system dependent
and generally gives rise to corner-cutting and
sliding-down problems~\cite{jonssonmj98}. It is convenient to discuss these 
difficulties in terms of the
components of the true gradient, $\gt$, and spring gradient, $\gs$, parallel and perpendicular to the path.
The parallel component of the gradient $\gtpar$ at 
image $i$ on the path is obtained by projecting out the perpendicular component
$\gtper$ using an estimate of the tangent to the path. 
The parallel and perpendicular components for image $i$ are:
\begin{equation}
    \gtpar_i = \left( \bflp_i V_i \cdot \bft_i \right) \bft_i, \hskip1cm \gtper_i = \bflp_i V_i - \gtpar_i,
    \label{eq:pp}
\end{equation}
where $V_i = V\left( \bfx_i \right)$, and the unit vector $\bft_i$ is the tangent. 
Here and throughout this paper we denote unit vectors by a hat.
The complete gradient, $\gt$, has $N \times \eta$ components for a band
of $N$ images with $\eta$ atomic degrees of freedom each.

Corner-cutting has a significant effect when a path experiences high curvature.
Here $\gsper$ is large, which prevents the images from closely following the path
because the spring force necessarily has a significant component perpendicular to the tangent. 
The sliding-down problem occurs due to the presence of $\gtpar$,
which perturbs the distribution of images along the path, creating high-resolution
regions (around the local minima) 
and low-resolution regions (near the transition states) \cite{jonssonmj98}. 
Both problems significantly affect the ability
of the NEB method to produce good transition state candidates.
We have found that
sliding-down and corner-cutting are interdependent and cannot
both be remedied by adjusting the spring force constant $k_{spr}$;
increasing $k_{spr}$ may prevent sliding-down but it will make corner-cutting worse. 

The aforementioned problems can sometimes be eliminated by constructing the NEB gradient 
from the potential in the following way:
$\gtpar$ and $\gsper$ are projected out, which
gives the elastic band its `nudged' property~\cite{henkelmanj00}. 
Removal of $\gtpar$ can be thought of as
bringing the path into a plane or flattening the 
PES [Figure \ref{fig:surface}(b)], while removal of $\gsper$ is
analogous to making the images heavier so that they favour the bottom of the valley at all times.

\begin{figure}
\centerline{\includegraphics{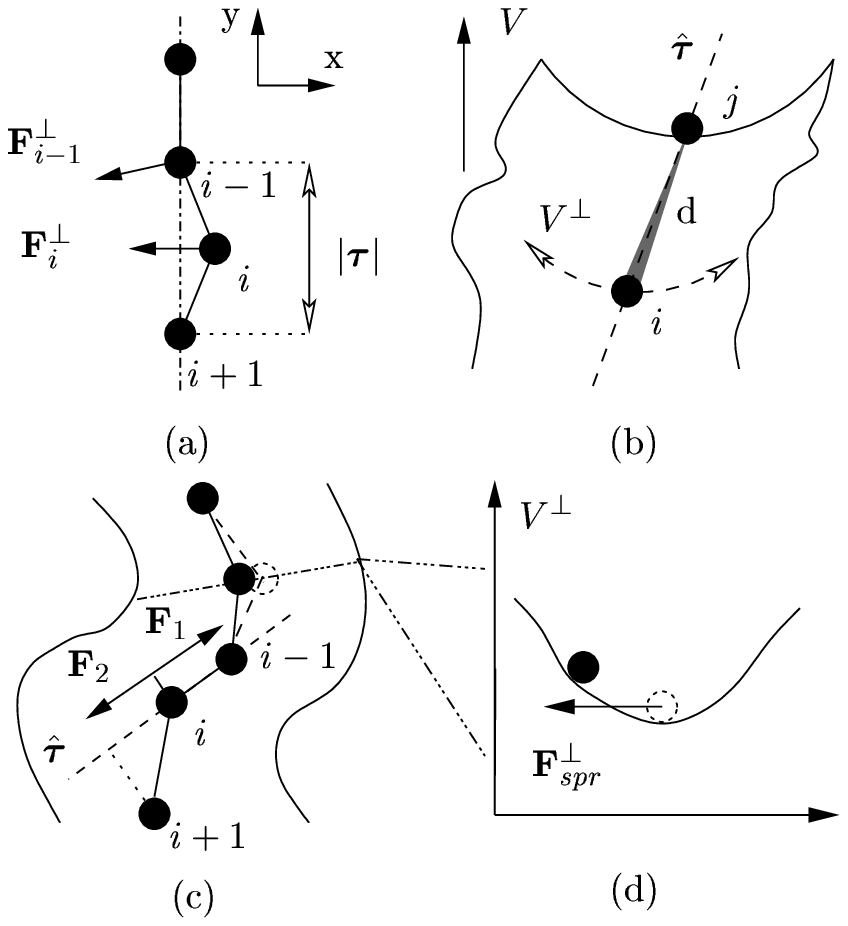}}
\caption{}
\label{fig:tech}
\end{figure}

The choice of a method to estimate the tangent to the path is important for it affects the convergence
of the NEB calculation. Originally, the tangent vector, $\bft_i$, for image $i$ was obtained by normalising
the line segment between the two adjacent images, $i+1$ and $i-1$~\cite{jonssonmj98}:
\begin{equation}
	\bft_i = \frac{ \bfx_{i+1}-\bfx_{i-1} }{\vert \bfx_{i+1}-\bfx_{i-1} \vert}.
	\label{eq:tauold}
\end{equation}
However, kinks can develop during optimisation of the image chain using this definition of $\bft_i$.
It has been shown~\cite{henkelmanuj00} that kinks are likely to appear in the regions where
the ratio $g_i^{\parallel}/g_j^{\perp}$ is
larger than the length of the line segment, $|\mbox{\boldmath$\tau$}|$, used in estimating the
tangent [Figure \ref{fig:tech}(a)].

Both the above ratio, the image density and $|\bft|$ can
vary depending on the system of interest, the particular pathway and other parameters of the NEB calculation.
From equation (\ref{eq:tauold})
it can be seen that $\bft_i$, and, hence, the next step in the
optimisation of image $i$, is determined by its neighbours, which are
not necessarily closer to the path than image $i$. 
Therefore, a better approach in estimating the $\bft_i$ would be to use only one
neighbour, since then we only need this neighbour to be 
better converged than image $i$.

There are two neighbours to select from, and it is natural 
to use the higher-energy one for this purpose, since steepest-descent paths are easier
to follow downhill than uphill:
\begin{equation}
	\bft_i = \frac{ \left( j-i \right) \left( \bfx_{j}-\bfx_{i} \right) }{\vert \bfx_{j}-\bfx_{i} \vert},
	\label{eq:taunew}
\end{equation}
where $i$ and $j$ are two adjacent images with energies $E_i$ and $E_j$, and $E_i < E_j$.
In this way, an image $i$ that has one higher-energy neighbour $j$ behaves as if it is 
`hanging' on to it [Figure \ref{fig:tech}(b)].

The above tangent formulation requires special handling of extrema along the path, and a
mechanism for switching $\bft$ at such points was
proposed~\cite{henkelmanuj00}. It also fails to produce an even distribution of images
in regions with high curvature [Figure \ref{fig:tech}(c)].
We presume that Henkelman and J\'onsson substitute 
$\left( \gs \cdot \bft \right) \bft$ by $\vert \gs \vert \bft$ 
in equation (\ref{eq:pp})
to obtain a spring gradient formulation that will keep
the images equispaced when the tangent from equation (\ref{eq:taunew}) is used in the
projections:~\cite{henkelmanj00}
\begin{equation}
	\gspar_i = k_{spr} \left( \vert \bfx_i - \bfx_{i-1} \vert - \vert \bfx_{i+1} - \bfx_i \vert \right) \bft_i.
	\label{eq:newgspr}
\end{equation}

We have previously used the NEB approach to produce candidate transition state guesses for
further refinement using hybrid EF methods \cite{wales02,evansw03},
which avoid either calculating the Hessian or diagonalising it \cite{munrow99,kumedamw01}.
Having obtained tightly converged transition states approximate steepest-descent paths are
calculated by energy minimisation, as discussed in \S \ref{sec:connect}.

In the present work
the NEB approach has been used in combination with two minimisers, namely
the quenched velocity Verlet (QVV) and the limited-memory
Broyden--Fletcher--Goldfarb--Shanno (L-BFGS) algorithms.
The QVV method is based on
the velocity Verlet algorithm~\cite{allent89} (VV) as modified by
J\'{o}nsson {\em et al.}~\cite{jonssonmj98} and was originally used for NEB optimisation.
VV is a symplectic integrator that enjoys widespread
popularity, primarily in molecular dynamics (MD) simulations where it is used
for numerical integration of Newton's equations of motion.
At each time step $\delta t$ the coordinates and the velocities $\mbox{\boldmath$\mathcal{V}$}$
are updated from the coupled first-order differential equations in the following manner~\cite{allent89}:
\begin{align}
     &\bfx \left(t+\delta t\right) =  \bfx \left(t\right) + \delta t \mbox{\boldmath$\mathcal{V}$} \left(t\right) -
     \frac{\delta t^2}{2 m} \gt \left(t \right) \label{eq:qvva} \\
     & \mbox{\boldmath$\mathcal{V}$}\left(t+\frac{1}{2} \delta t \right) = \mbox{\boldmath$\mathcal{V}$}\left(t \right) -
     \frac{\delta t}{2 m} \gt \left(t\right) \label{eq:qvvb} \\
     & \mbox{\boldmath$\mathcal{V}$}\left(t+ \delta t \right) = \mbox{\boldmath$\mathcal{V}$}\left(t+\frac{1}{2} \delta t
     \right) - \frac{\delta t}{2 m} \gt \left(t+ \delta t\right) \label{eq:qvvc}
\end{align}
The algorithm involves two stages, with a force evaluation in between. First the positions
are updated according to equation
(\ref{eq:qvva}), and the velocities at midstep $t + \delta t/2$ are then computed using
equation (\ref{eq:qvvb}).
After the evaluation of the gradient at time $t + \delta t$ the velocity is updated again
[equation (\ref{eq:qvvc})] to complete the move.
To obtain minimisation it is necessary to remove kinetic energy, and this can be done in
several ways. If kinetic energy is removed completely every step the algorithm is
equivalent to a steepest-descent minimisation, which is rather inefficient.
Instead, it was proposed by J\'{o}nsson {\em et al.}~\cite{jonssonmj98} to keep only the velocity 
component that is antiparallel to the gradient
at the current step. If the force is consistently pointing in the
same direction the system accelerates, which is equivalent to increasing the time
step~\cite{jonssonmj98}. However, a straightforward variable time step
version of the above algorithm was reported to be unsuccessful~\cite{crehuetf03}.  

L-BFGS is a version of the BFGS algorithm that limits the
storage used and is hence particularly suitable for
large-scale problems~\cite{liun89}. The difference between the L-BFGS algorithm and
standard BFGS is in the Hessian matrix update.
In order to store each correction to
the Hessian $2n$ storage locations are needed, 
$n$ being the dimensionality of the problem~\cite{nocedal80}. L-BFGS stores a maximum of $m$
corrections and the Hessian matrix is never formed explicitly.
Every iteration $-\hess^{-1}{\bf g}$ is computed according to a
recursive formula described by Nocedal~\cite{nocedal80}. For the first $m$ iterations
L-BFGS is identical to the BFGS method, but after these the oldest correction is discarded.
Since only the $m$ most recent corrections are retained L-BFGS uses less storage.
Here we employed a modified version of Nocedal's L-BFGS implementation~\cite{lbfgs} in which 
the line search was removed and the maximum step size was limited for each image separately.

It is noteworthy that the objective function corresponding to the projected NEB gradient is unknown,
but it is not actually required in either of the minimisation algorithms that we consider.

\section{Results}
\label{sec:results}
The springs should distribute the images evenly along the NEB path during the
optimisation, and the choice of $k_{spr}$ must be made at the beginning of each run. 
It has been suggested by J\'{o}nsson and coworkers that since the action of the springs is only felt along
the path the value of the spring constant is not critical as long as it is not zero
\cite{jonssonmj98}. If
$k_{spr}$ is set to zero then convergence is guaranteed for the first several tens of iterations 
only; even though $\gtpar$ is
projected out, $\bft$ fluctuates and further optimisation will eventually result in the majority of 
images gradually sliding down to local minima~\cite{jonssonmj98}.

\subsection{Slow-response Quenched Velocity Verlet}
\label{sec:qvv}
In practice we find that the value of $k_{spr}$ affects the convergence properties and 
the stability of the optimisation process.
This result depends on the type of minimiser employed and may also depend on minimiser-specific settings.
Here we analyse the convergence properties of NEB minimisations using the QVV minimiser
(NEB/QVV) and their dependence on the type of velocity quenching.
From previous work it is not clear when is the best
time to perform quenching during the MD minimisation of the NEB
\cite{jonssonmj98,henkelmanuj00,henkelmanj00}. Since the VV algorithm calculates
velocities based on the gradients at both current and previous steps
quenching could be applied using either of these gradients.

\begin{figure}
\centerline{\includegraphics{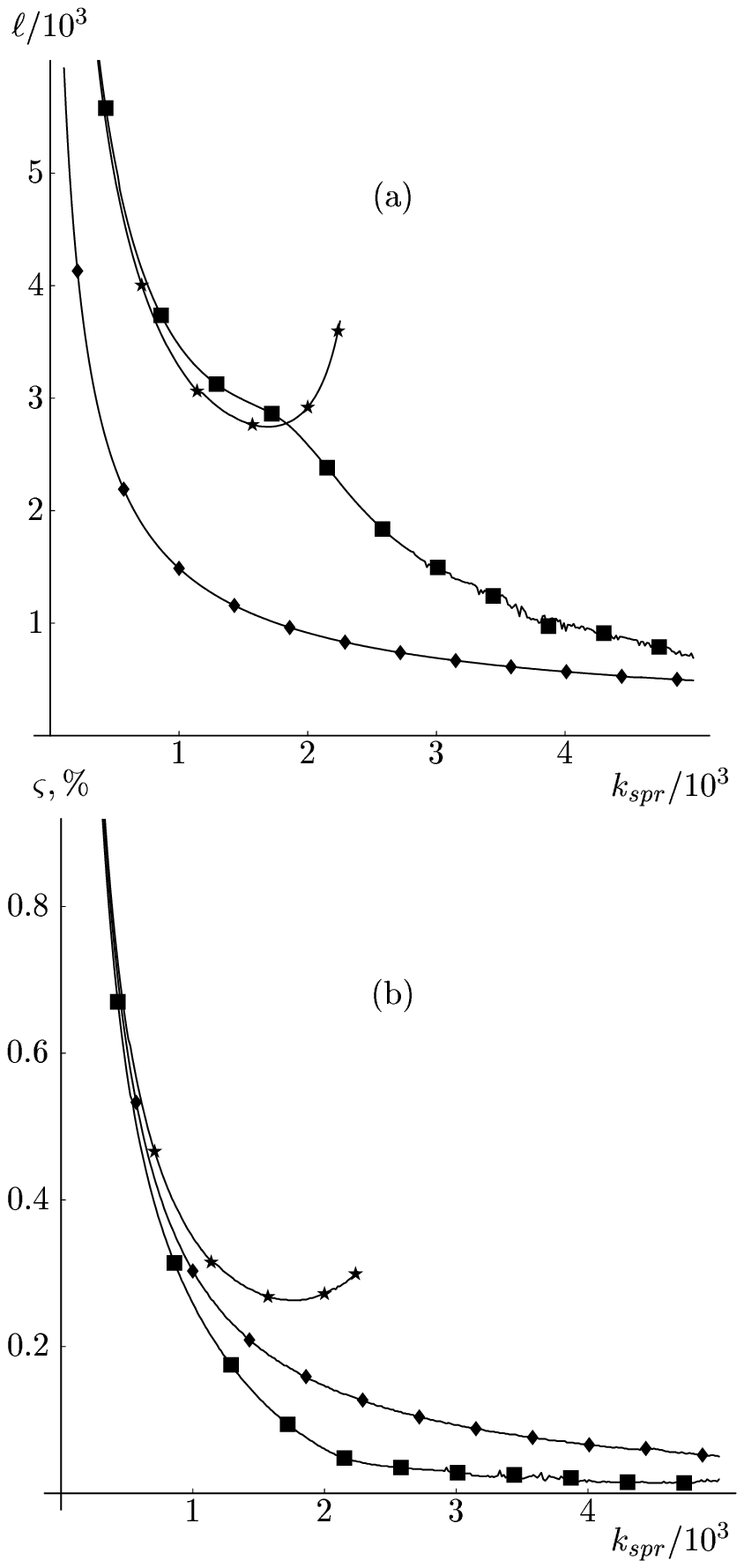}}
\caption{}
\label{fig:q}
\end{figure}

Specifically, it is possible to quench velocities right after advancing the system using
equation (\ref{eq:qvva}), 
at the half-step in the velocity evaluation (quenching intermediate velocities at time $t + \delta t/2$)
using either the old or new gradient 
[equation (\ref{eq:qvvb})], or after completion of the velocity update. 
In Figure \ref{fig:q} we present 
results for the stability of NEB/QVV as a function of the force constant parameter
for three of these quenching approaches.
We will refer to an NEB optimisation as stable for a certain
combination of parameters (e.g. time integration step, number of images)
if the NEB steadily converges to a well-defined path and/or stays in its proximity until
the maximal number of iterations is reached or the convergence criterion is satisfied.

Figure \ref{fig:q} shows the results of several thousand optimisations
for a 17-image band with the M{\"{u}}ller-Brown two-dimensional
potential~\cite{mullerb79} using QVV minimisation and
a time step of $0.01$ (consistent units) for different values of $k_{spr}$.
This widely used surface does not present a very challenging or realistic test case, but if an algorithm
does not behave well for this system it is unlikely to be useful.
Each run was started from the initial guess obtained using linear interpolation and terminated when
the root-mean-square (RMS) gradient became less than $0.01$. We define the RMS
gradient for the NEB as
\begin{equation}
     g^{\perp}_{\text{RMS}} = \sqrt{ \frac{ \sum_{i=1}^{N} \lvert \gtper_i \rvert}{N \eta} }
	\label{grms}
\end{equation} 
where $N$ is the number of images in the band 
and $\eta$ is the number of atomic degrees of freedom available to each image.

It seems natural to remove the velocity component perpendicular to
the gradient at the current point when the geometry $\bfx \left(t\right)$, gradient
$\gt \left(t\right)$ and velocity $\mbox{\boldmath$\mathcal{V}$}\left(t\right)$ are
available, i.e.
\begin{equation}
     \mbox{\boldmath$\mathcal{V}$}_{\text{Q}}\left(t\right) =
	\Bigl(\mbox{\boldmath$\mathcal{V}$}\left(t\right) \cdot \hat{\gt}\left(t\right)\Bigr)
        \hat\gt\left(t\right),
	\label{eq:sqvv}
\end{equation}
where $\mbox{\boldmath$\mathcal{V}$}_{\text{Q}}\left(t\right)$ is the velocity vector after quenching.
However, we found this approach to be the least stable of all---the optimisation was slow and
convergence was very sensitive to the magnitude of time step. Hence we do not show any
results for this type of quenching.

From Figure \ref{fig:q}(a) we see that the best approach 
is to quench the velocity after the coordinate update.
The optimisation is then stable for a wide range of force constant values, 
and the images on the resulting pathway are evenly distributed.
In this quenching formulation the velocity response to a new gradient
direction is retarded by one step in coordinate space: the step is still taken in the direction
$\mbox{\boldmath$\mathcal{V}$}\left(t\right)$ but the corresponding velocity component is 
removed. 
To implement this slow-response QVV (SQVV) it is necessary to modify the VV algorithm described in section
\ref{sec:qvv} by inserting equation 
(\ref{eq:sqvv}) in between the two stages described by equations (\ref{eq:qvva}) and
(\ref{eq:qvvb}).

The second best approach after SQVV is to quench the velocity at
midstep $t+\delta t/2$  using the new gradient. 
On average, this algorithm takes twice as long to converge the NEB to a given RMS gradient
tolerance compared to SQVV. However, the method is stable for the same range of spring force
constant values and produces a pathway in which the images are equispaced more accurately than the
other formulations [see Figure \ref{fig:q}(b)].

The least successful of the three QVV schemes considered
involves quenching velocities at mid-step using the gradient
from the previous iteration (stars in Figure \ref{fig:q}). Even though
the number of iterations required is roughly comparable to that obtained by quenching 
using the new gradient, it has the smallest
range of values for the force constant where it is stable.
Some current implementations of NEB~\cite{alfonsoj03,alfonso} 
(intended for use in combination with electronic structure
codes) use this type of quenching in their QVV implementation.

We have also conducted analogous calculations for more complicated systems such as 
permutational rearrangements of Lennard-Jones clusters.
The results are omitted for brevity, but agree with the conclusions drawn from the 
simpler 2D model described above. The same is true for the choice of force constant
investigated in the following section.

\subsection{Choice of the Force Constant}
\label{sec:k}
We find that
if the force constant is too small many more iterations are needed to converge the images
to the required RMS tolerance, regardless of the type of quenching. In addition,
the path exhibits a more uneven image distribution.
This result occurs because at the initial stage
the images may have very different gradients from the true potential
along the band, because they lie far from  the required path,
and the true potential gradient governs the optimisation. 
When the true RMS force is reduced the springs start
to play a more important role. But at this stage the forces are small and so is the QVV step size.
The influence of the springs is actually most important during the initial optimisation stage, 
for it can determine the placement of images in appropriate regions.
It is less computationally expensive to guide an image into the right region at
the beginning of an optimisation
than to restore the distribution afterwards by dragging it between two minima through a 
transition state region.

If $k_{spr}$ is too big the NEB never converges to the required RMS gradient tolerance value.
Instead, it stays in proximity to the path but develops oscillations: adjacent images 
start to move in opposite directions.
For all types of quenching we observed similar behaviour when large values of the force constant were used.
This problem is related to the step in coordinate space that the optimiser is taking: for 
the SQVV case simply decreasing the time step remedies this problem.

\subsection{Comparison of SQVV and L-BFGS Minimisers for the MB Surface}
We tested the NEB/L-BFGS method by minimising a 17-image NEB for the two-dimensional
M{\"{u}}ller-Brown surface~\cite{mullerb79}. 
Our calculations
were carried out using the OPTIM program~\cite{optim03}. The NEB method in its previous
formulation~\cite{henkelmanj00} and a modified L-BFGS minimiser~\cite{wales03}
were implemented in OPTIM in a previous discrete path sampling study~\cite{wales02}. We used the same
number of images, initial guess and termination criteria as described in section
\ref{sec:qvv} to make the results directly comparable.

\begin{figure}
\centerline{\includegraphics{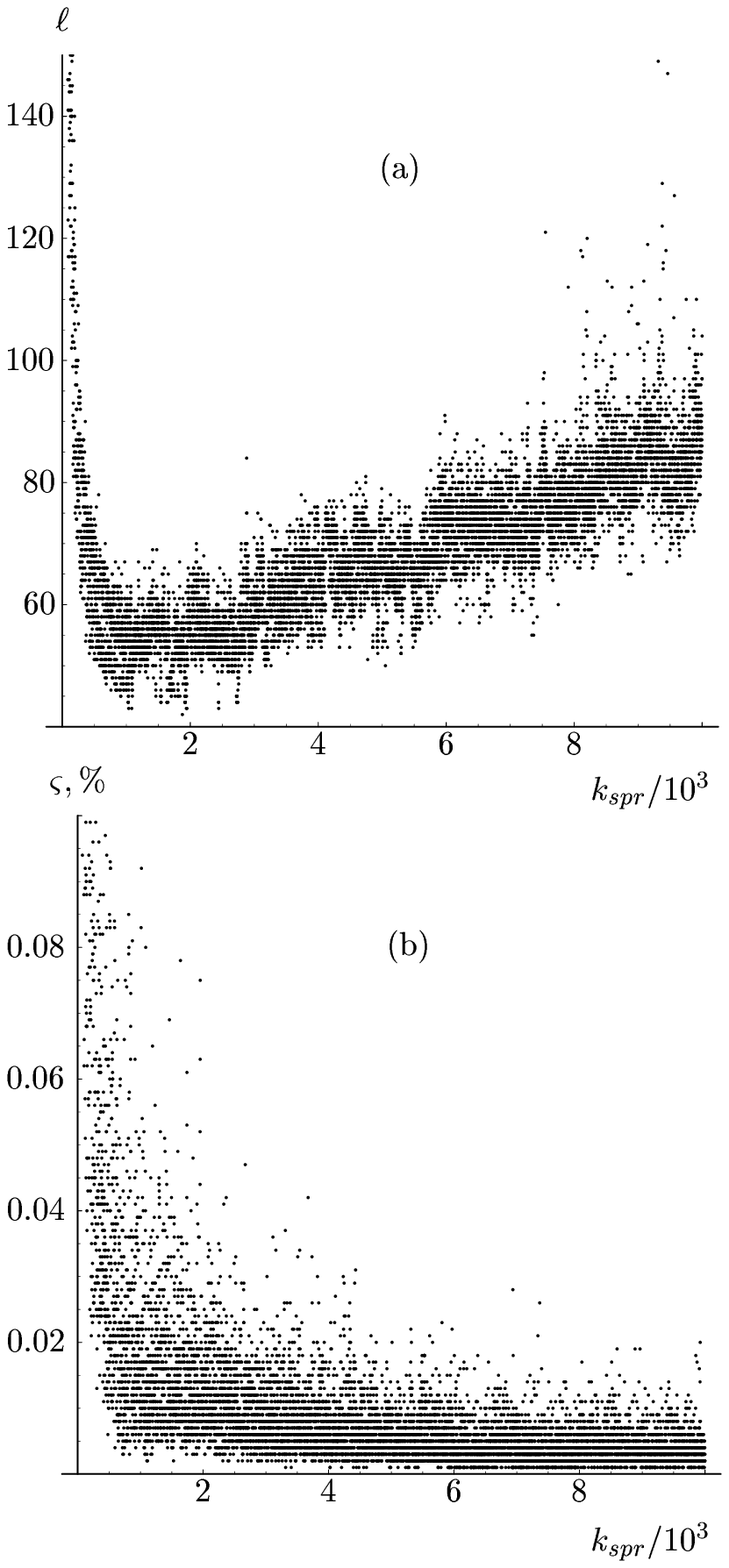}}
\caption{}
\label{fig:lbfgsMB}
\end{figure}

Figure \ref{fig:lbfgsMB} shows the performance of the L-BFGS minimiser as
a function of $k_{spr}$.
We used the following additional L-BFGS specific settings. The number of corrections in
the BFGS update was set to $m=4$ (Nocedal's recommendation for the number of corrections is
$3 \leqslant m \leqslant 7$~\cite{lbfgs}), the maximum step size 
was $0.1$, and we limited the step size for each image separately, i.e.
\begin{equation}
     \lvert {\bf p}_j \rvert \leqslant 0.1,
\end{equation}
where ${\bf p}_j$ is the step for image $j$. The diagonal elements of the inverse
Hessian  were initially set to $0.1$.

From Figure \ref{fig:lbfgsMB} it can be seen that the performance of
L-BFGS minimisation is relatively independent
of the choice of force constant. All the optimisations with $30 \leqslant k_{spr} \leqslant 10,000$
converged to the steepest-descent path, and, for most of this range, in less than 100
iterations. This method therefore gives roughly an order of magnitude improvement in speed over SQVV
minimisation [see Figure \ref{fig:q}(a)].

We found it helpful to limit the step size while optimising the NEB with the L-BFGS minimiser.
The magnitude and direction of the gradient on adjacent images can vary significantly.
Taking bigger steps can cause the appearance of temporary discontinuities and kinks in the NEB.
The NEB still converges to the correct path, but it takes a while for these
features to disappear and the algorithm does not converge any faster.

\subsection{Doubly Nudged Elastic Bands}
The NEB/QVV approach
has previously been systematically tested on systems with around 100 degrees of
freedom, $\eta$~\cite{maragakisabrk02}. However, in the majority of cases these
test systems could be divided into a `core' and a smaller part that actually changes significantly.
The number of active degrees of freedom
is therefore significantly smaller than the total number in these tests.
For example, prototropic tautomerisation of cytosine nucleic
acid base ($\eta = 33$) involves motion of one hydrogen atom along a quasi-rectilinear
trajectory accompanied by a much smaller distortion of the core.

We have therefore tested the performance of the NEB/SQVV and NEB/L-BFGS schemes for more complicated
rearrangements of Lennard-Jones (LJ) clusters to validate the results of \S \ref{sec:k},
and to investigate the stability and performance of both approaches when there are more active
degrees of freedom.
Most of our test cases involve permutational isomerisation of the LJ$_7$, 
LJ$_{38}$ and LJ$_{75}$ clusters.
These examples include cases with widely varying separation between the endpoints, 
integrated path length, number of active degrees of freedom and cooperativity.

Permutational rearrangements are particularly interesting because it is relatively
difficult to produce an initial guess for the NEB run.
In contrast, linear interpolation between the endpoints was found to provide a useful
initial guess for a number of simpler cases~\cite{jonssonmj98}. 
For example, it was successfully used to construct the NEB for rearrangements that involve one
or two atoms following approximately rectilinear
trajectories, and for migration of a single atom on a surface~\cite{maragakisabrk02}. 
For more complex processes an alternative approach adopted in previous work is simply
to supply a better initial guess `by hand',
e.g. construct it from the images with unrelaxed geometries containing no 
atom overlaps~\cite{maragakisabrk02}.
The `detour' algorithm described in previous calculations that employ the ridge method could
also be used to avoid `atom-crashing' in the initial interpolation \cite{ionovac93}.

It has previously been suggested that it is important to
eliminate overall rotation and translation (ORT) of each image during 
the optimisation of an NEB~\cite{jonssonmj98}.
We have implemented this constraint in the same way as J\'onsson {\it et al.\/},
by freezing one atom, restricting the motion of a second atom to a plane,
and constraining the motion of a third atom to a line by zeroing the appropriate components of
the NEB gradient.

\begin{figure}
\centerline{\includegraphics{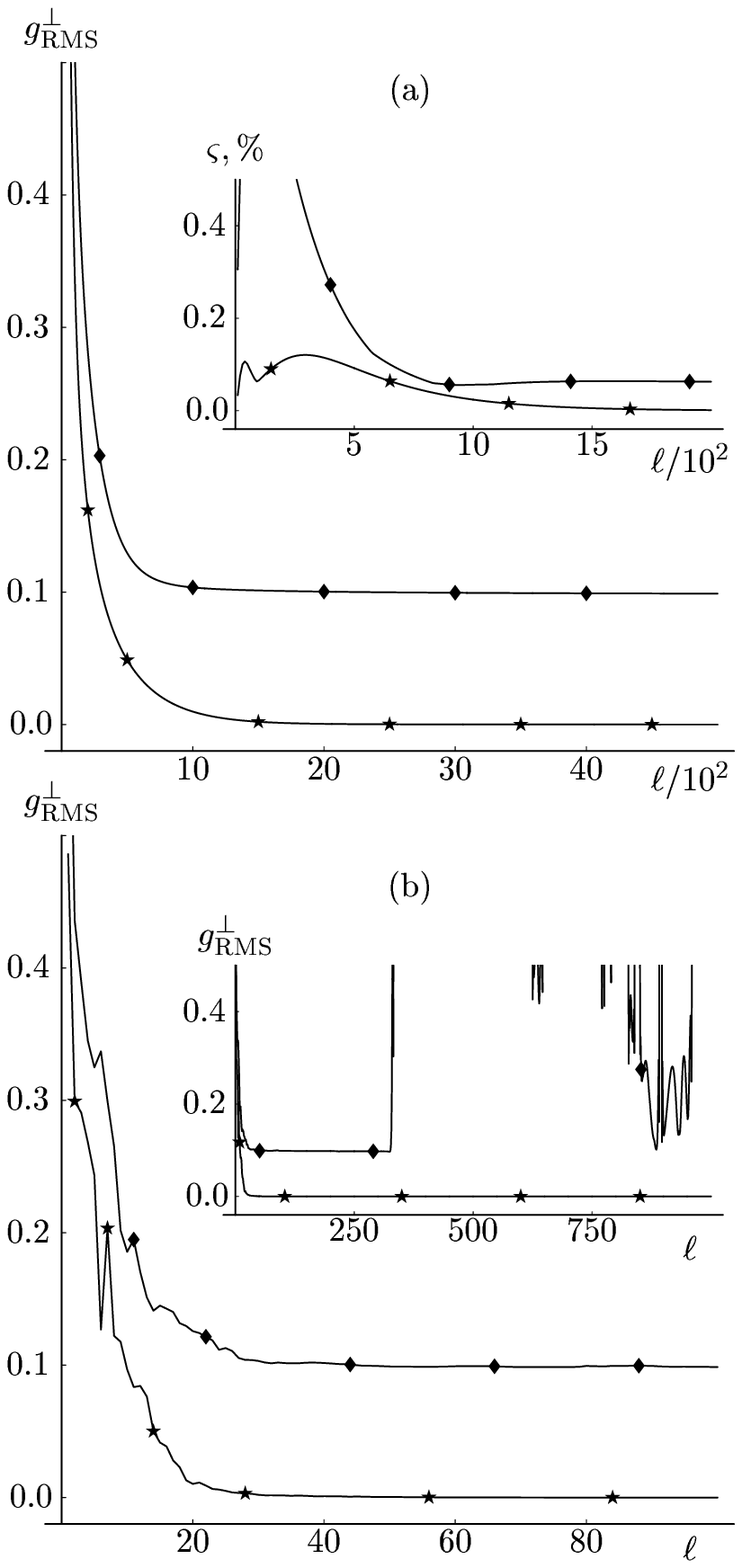}}
\caption{}
\label{fig:rottransl}
\end{figure}

We were able to obtain stable convergence in NEB/L-BFGS calculations only for
simple rearrangements, which confirms 
that straightforward L-BFGS optimisation of the NEB is unstable 
\cite{maragakisabrk02}. Figure \ref{fig:rottransl}
shows the performance of the 
NEB/SQVV [Figure \ref{fig:rottransl}(a)] and NEB/L-BFGS [Figure \ref{fig:rottransl}(b)]
approaches for one such rearrangement.
These calculations were carried out using a 7-image NEB both with (diamonds) and without (stars)
removing ORT for isomerisation 
of an LJ$_7$ cluster (global minimum $\rightarrow$ second-lowest minimum). The number of
iterations, $\ell$, is proportional to the number
of gradient evaluations regardless of the type of minimiser. 
Hence, from Figure \ref{fig:rottransl} we conclude
that for this system NEB/L-BFGS is faster than NEB/SQVV by approximately two 
orders of magnitude. 
However, removal of ORT leads to instability in the NEB/L-BFGS optimisation: 
the images do not stay in proximity to the required path
for long and instead diverge from it [see inset in Figure \ref{fig:rottransl}(b)].

By experimentation we have found that the main source of the instabilities is 
the complete removal of $\gsper$.
Instead, the inclusion of some portion of $\gsper$ in the NEB gradient, i.e.
\begin{equation}
     \gt_{\rm NEB} = \gtper + \gspar + \gs^{*},
	\label{eq:grad1}
\end{equation}
where $\gs^{*} = \xi \gsper$, makes the NEB/L-BFGS calculations stable but introduces some additional corner-cutting,
as well as an extra parameter, $\xi$. Since we use the transition state candidates
from NEB as starting points for 
further EF calculations
the corner-cutting is not a drawback as long as the transition state candidates are good
enough. By adjusting $\xi$ in the range of $\left(0.01,0.1\right)$ we were able to
achieve satisfactory performance for the NEB/L-BFGS method in a number of cases.
However, an alternative modification, described below, proved to be even more
successful.

The drawback of the NEB gradient described by equation (\ref{eq:grad1}) stems from the interference of
$\gtper$ and $\xi \gsper$, and becomes particularly noticeable when the projection 
of $\xi \gsper$ on $\gtper$
and $\gtper$ itself are of comparable magnitude. This problem is analogous to the interference of
$\gt$ and $\gs$ in the original elastic band method, which was previously solved by `nudging'~\cite{henkelmanj00}. 
We have therefore constructed
the gradient of a new `doubly' nudged elastic band (DNEB) 
using
\begin{equation}
 	\gs^{*} = \gsper - (\gsper \cdot \hgtper)\hgtper.
	\label{eq:double}
\end{equation}
In this formulation some corner-cutting may still occur because 
the images tend to move cooperatively during optimisation;
the spring gradient $\gs^{\perp}_{\rm DNEB}$ acting on one image
can still indirectly interfere with the true gradients of its neighbours.
In our calculations this drawback was not an issue, since
we are not interested in estimating properties of the path directly from
its discrete representation. Instead we construct it from steepest-descent paths
calculated after converging the transition states tightly using the EF approach.
We have found DNEB perfectly adequate for this purpose.

We have also tested a number of approaches that might be useful if one
wants to produce a full pathway involving a number of
transition states for a complicated rearrangement in just one NEB run.
One of these, for instance, is a gradual removal of the $\gs^{*}$
component from the NEB gradient once some convergence criterion is achieved. 
This removal works remarkably well,
particularly in situations with high energy initial guesses, which occur frequently if the guessing is fully
automated. This adjustment can be thought of as making the band less elastic in the beginning in order to resolve
the highest-energy transition state regions first.

\subsection{Comparison of the DNEB/L-BFGS and DNEB/SQVV Methods for 
Permutational Isomerisations of LJ$_7$}
It is sometimes 
hard to make a direct comparison of different double-ended methods for a particular rearrangement
because the calculations may converge to different paths.
Another problem concerns the choice of a consistent termination criterion: 
the RMS force usually converges to some finite system-dependent value,
which in turn may depend on the number of images and other parameters.
A low-energy chain of NEB images does not necessarily mean that a good pathway has been obtained,
since it may arise because more images are associated with regions around local minima,
rather than the higher energy transition state regions.
Here we present the results of DNEB/L-BFGS, DNEB/SQVV and, where possible, NEB/SQVV calculations for
all the distinct permutational rearrangements of the global minimum 
for the LJ$_7$ cluster (see Figure~\ref{fig:ep} for the endpoints and nomenclature).

It is possible to draw a firm conclusion as to how well the NEB represents 
the pathway when the corresponding stationary points and steepest-descent paths are already known.
We therefore base our criterion for the effectiveness of an NEB calculation 
on whether we obtain good estimates of all the
transition states. By considering several systems
of increasing complexity we hope to obtain comparisons that are not specific to a particular pathway.

Connections between two minima are defined by calculating an approximation to the two 
steepest-descent paths that lead downhill from each transition state,
and two transition states are considered connected if they are
linked to the same minimum via a steepest-descent path.
We will say that minima are `connected' if there exists a path consisting of one or more
transition states and intermediate minima linking them.
Permutational isomers of the same minimum are distinguished in these calculations.
We refer to the chain of images
produced by the NEB calculation as `connected' if going downhill from each transition state
using steepest-descent minimisation yields a set of minima that contains the endpoints linked together.

For NEB/SQVV calculations we used the NEB 
formulation defined in Ref.~\onlinecite{henkelmanj00}.
DNEB is different from the above method because it includes an
additional component in the NEB gradient, as described by equations (\ref{eq:grad1}) and
(\ref{eq:double}). In addition, for the following DNEB calculations
we did not remove overall rotation and translation (ORT), 
because we believe it is unnecessary when our gradient modification is used.
To converge transition state candidates tightly we employed EF optimisation, 
limiting the maximum number of EF iterations to five with
an RMS force convergence tolerance of $10^{-5}$. 
(Standard reduced units for the Lennard-Jones potential are used throughout this paper.)
Initial guesses for all the following calculations were obtained by
linear interpolation between the endpoints. 
To prevent `atom-crashing' from causing overflow in the initial guess we simply perturbed
such images slightly using random atomic displacements of order $10^{-2}$ reduced units.

In each case we first minimised the Euclidean distance between the endpoints 
with respect to overall rotation and translation
using the method described in Ref.~\onlinecite{rhee00}.
SQVV minimisation was performed with a time step of $0.01$ and a maximum step size per 
degree of freedom of $0.01$. This limit on the step size
prevents the band from becoming `discontinuous' initially and 
plays an important role only during the first $100$ or so iterations.
The limit was necessary because for the cases when the endpoints are
permutational isomers linear interpolation usually yields bands with 
large gradients, and it is better to refrain from taking excessive steps at this stage. 
We did not try to select low energy initial guesses for each rearrangement individually,
since one of our primary concerns was to automate this process. 
For the same reason, all the L-BFGS optimisations were started from
guesses preoptimised using SQVV until the RMS force dropped below $2.0$.

Table \ref{tab:1} shows the minimum number of images and gradient calls 
required to produce a connected pathway using the
DNEB/L-BFGS and DNEB/SQVV methods. These calculations were run assuming no prior knowledge of the path.
Normally there is no initial information available on the integrated path length or the number of
intermediate minima between the endpoints, 
and it takes some experimentation to select an appropriate number of images.
Our strategy is therefore to gradually increase the number of images to 
make the problem as computationally inexpensive as possible. 
Hence we increment the number of images and maximum number of 
NEB iterations in each calculation until a connected path is produced, in the sense 
defined above.
The permitted image range was $2\le N\le20$ 
and the maximum number of NEB iterations ranged from $1 \le l \le 3,000N$.
We were unable to obtain connected pathways for any of the four LJ$_7$ rearrangements 
using the NEB/SQVV approach.

\begin{figure}
\centerline{\includegraphics{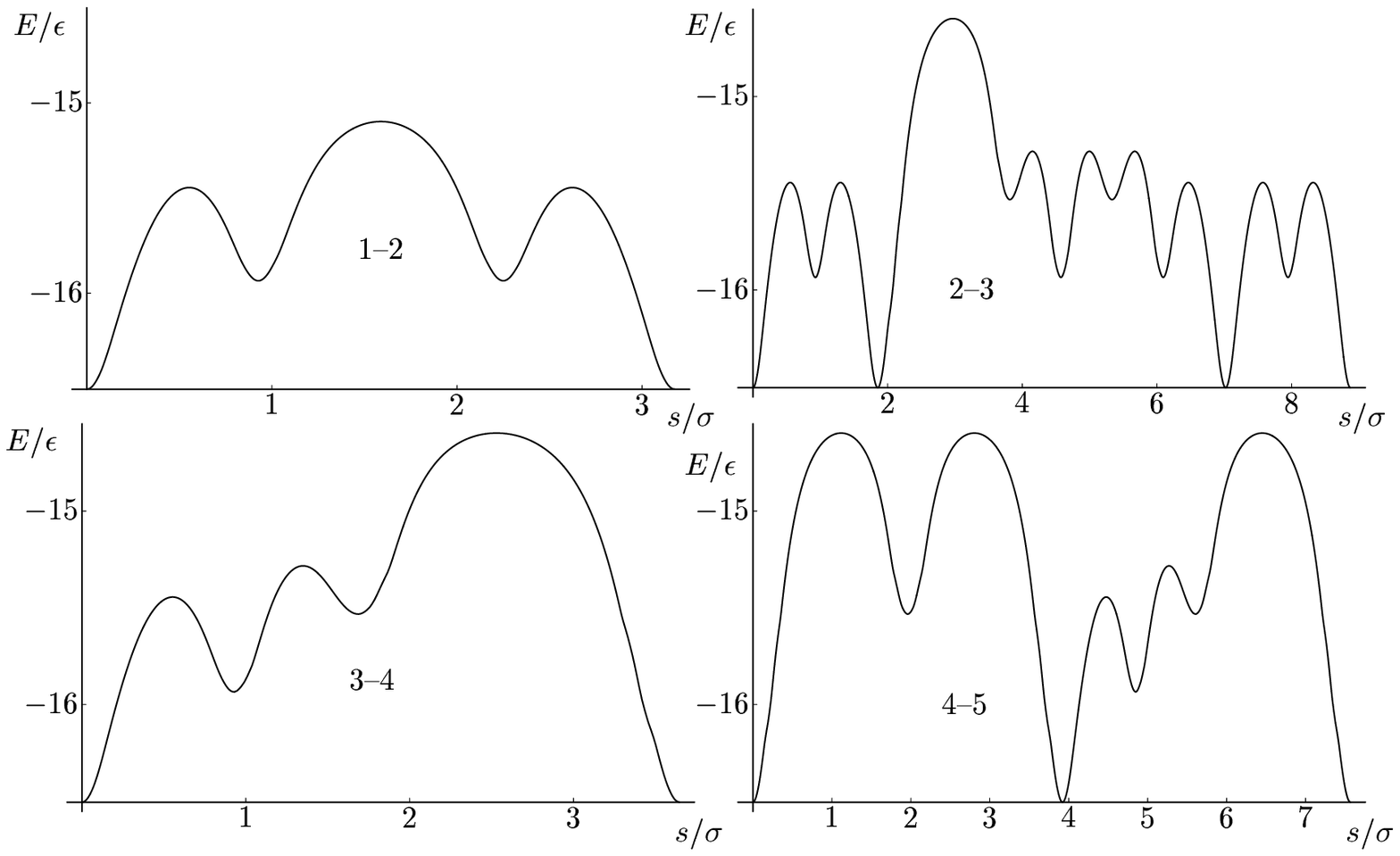}}
\caption{}
\label{fig:profiles}
\end{figure}

Table \ref{tab:2} presents the results of analogous calculations where 
we keep the number of images fixed to 50. Unlike the performance
comparison where the number of images is kept to a minimum (Table \ref{tab:1}),
these results should provide insight into the performance of the DNEB approach
when there are sufficient images to resolve all the transition states. 
All the optimisations for a particular rearrangement
converged to the same or an enantiomeric pathway unless stated otherwise. 
The energy profiles that correspond to these rearrangements
are shown in Figure \ref{fig:profiles}.

From Table \ref{tab:1} and \ref{tab:2} we conclude that in all cases the
DNEB/L-BFGS approach is more than an order of magnitude faster than DNEB/SQVV.
It is also noteworthy that the DNEB/SQVV approach is faster than NEB/SQVV
because overall rotation and translation are not removed.
Allowing the images to rotate or translate freely can lead to numerical problems, namely a
vanishing norm for the tangent vector,
when the image density is very large or the spring force constant is too small.
However, when overall rotation and translation are not allowed there is less scope for 
improving a bad initial guess, because the images are more constrained.
This constraint usually means that more images are needed
or a better initial guess is required.
Our experience is that such constraints usually slow down convergence, 
depending on which degrees of freedom are frozen:
if these are active degrees of freedom (see above) the whole cluster must move instead,
which is usually a slow, concerted multi-step process. 

\subsection{A revised connection algorithm}
\label{sec:connect}
In previous work we have used the NEB approach to supply transition state guesses for further EF
refinement \cite{wales02,evansw03}. 
Double-ended searches are needed in these discrete path 
sampling runs to produce alternative
minimum--transition state--minimum$\cdots$ sequences from an initial path. 
The end minima that must be linked in such calculations
may be separated by relatively large distances,
and a detailed algorithm was described for building up a connected path using successive transition state
searches. 
The performance of the DNEB/L-BFGS approach is sufficiently good that we have changed this connection strategy
in our OPTIM program. In particular, the DNEB/L-BFGS method can often provide good candidates
for more than one transition state at a time, and may even produce all the necessary transition states on a
long path. However, it is still generally necessary to consider multiple searches
between different minima in order to connect a pair of endpoints. In particular, we would like to use the
minimum number of NEB images possible for reasons of efficiency, 
but automate the procedure so that it eventually succeeds or
gives up after an appropriate effort for any pair of minima that may arise in a discrete path sampling run.
These calculations may involve the construction of many thousands of discrete paths. 
As in previous work
we converge the NEB transition state candidates using eigenvector-following techniques 
and then use L-BFGS energy
minimisation to calculate approximate steepest-descent paths. 
These paths usually lead to local minima, which we also
converge tightly. The combination of NEB and hybrid eigenvector-following techniques 
\cite{munrow99,kumedamw01} is similar to
using NEB with a `climbing image' as described in Ref.~\onlinecite{henkelmanuj00}

The initial parameters assigned to
each DNEB run are the number of images and the number of iterations,
which we specify by image and iteration densities. 
The iteration density is the maximum number of iterations per image, while the image density
is the maximum number of images per unit distance. The distance in 
question is the Euclidean separation of the
endpoints, which provides a crude estimation of the integrated path length.
This approach is based on the idea that knowing the integrated path length, which
means knowing the answer before we start, 
we could have initiated each DNEB run with the same number of images per
unit of distance along the path. 
In general it is also impossible to provide a lower bound on the number of images necessary
to fully resolve the path, since this would require prior knowledge of the number of
intervening stationary points.
Our experience suggests that a good strategy is to employ as small an image and iteration density
as possible at the start of a run, and only increase these parameters for connections
that fail. 

All NEB images, $i$, for which $E_i>E_{i\pm 1}$ are considered
for further EF refinement. The resulting distinct
transition states are stored in a database and the corresponding energy minimised paths were
used to identify the minima that they connect. 
New minima are also stored in a
database, while for known minima new connections are recorded.
Consecutive DNEB runs aim to build up a connected path by progressively filling in connections
between the endpoints or intermediate minima to which they are connected.
This is an advantageous strategy
because the linear interpolation guesses usually become better as the separation
decreases, and therefore fewer optimisation steps are required.
Working with sections of a long path one at a time
is beneficial because it allows the algorithm to increase the
resolution only where it is needed.
Our experience is that this approach is generally significantly faster than trying to 
characterise the whole of a complex path with a single chain of images.

When an overall path is built up using successive DNEB searches we must select 
the two endpoints for each new search from the database of known minima.
It is possible to base this choice on the order in which the transition states were found,
which is basically the strategy used in our previous work \cite{wales02,evansw03}.
However, when combined with the new DNEB approach a better strategy is 
to connect minima based upon their Euclidean separation.
For this purpose it is convenient to classify all the minima
into those already connected to the starting endpoint (the S set),
the final endpoint (the F set), and the remaining minima, which are not connected to either endpoint
(the U set). 
The endpoints for the next DNEB search are then chosen as the
two that are separated by the shortest distance, where one belongs to
S or F, and the other belongs to a different set.
The distance between these endpoints is then minimised with respect to overall rotation and
translation, and an initial guess for the image positions is obtained 
using linear interpolation.
Further details of the implementation of this algorithm and the 
OPTIM program are available from authors upon request.

\begin{figure}
\centerline{\includegraphics{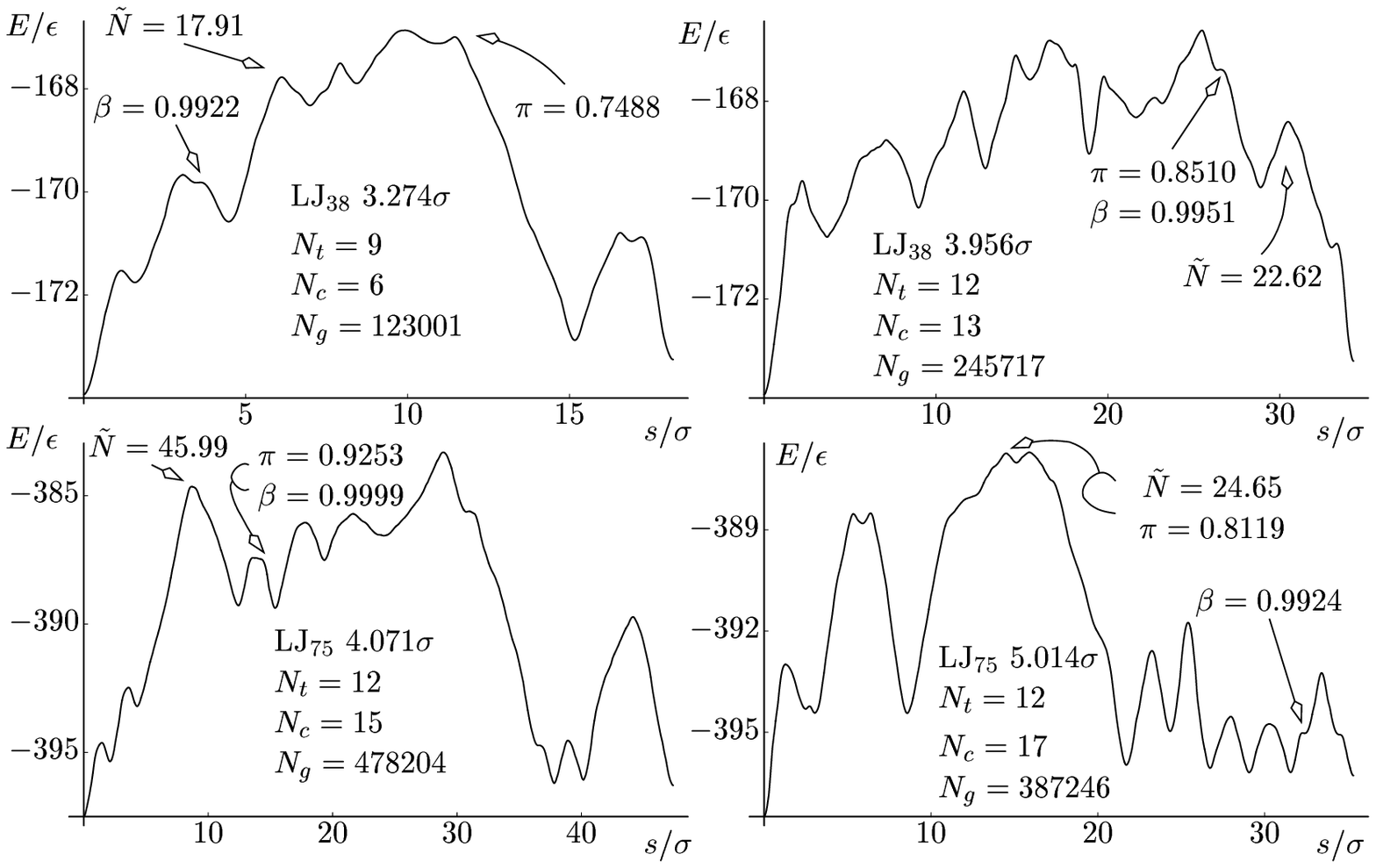}}
\caption{}
\label{fig:tests}
\end{figure}

As test cases for this algorithm we have considered
various degenerate rearrangements of LJ$_7$, LJ$_{13}$, LJ$_{38}$ and LJ$_{75}$. 
(A degenerate rearrangement is one that links permutational isomers of the same structure
\cite{leones70,wales03}.)
The PES's of LJ$_{38}$ and  LJ$_{75}$ have been analysed in a number of previous studies
\cite{doyemw99,millerdw99b,NeirottiCFD00,CalvoNFD00},
and are known to exhibit a double-funnel morphology: for both clusters
the two lowest-energy minima are structurally distinct and well separated in configuration space.
This makes them useful benchmarks for the above connection algorithm. Figure \ref{fig:tests}
depicts the energy profiles obtained using the revised connection algorithm for rearrangements between the
two lowest minima of each cluster. 
In each case we have considered two distinct paths that link different permutational
isomers of the minima in question, and these were chosen to be 
the permutations that give the shortest Euclidean distances.
These paths will be identified using the distance between the
two endpoints; for example, in the case of LJ$_{38}$ we have paths LJ$_{38}$ $3.274\,\sigma$
and  LJ$_{38}$ $3.956\,\sigma$, where $2^{1/6}\sigma$ is the pair equilibrium separation for the
LJ potential.

For each calculation we used the following settings: 
the initial image density was set to 10 and the iteration density to 30. 
If a connection failed for a particular pair of minima then up to two more attempts were
allowed before moving to the pair with the next smallest separation.
For the second and third attempts the number of images was increased by $50\%$ each time.
The maximum number of EF optimisation steps was set to $30$ with RMS force convergence
criterion of $10^{-5}$. In Figure \ref{fig:tests} every panel is labelled with 
the separation between the endpoints, the number 
of transition states in the final pathway, the number of DNEB runs required,
and the total number of gradient calls.

Individual pathways involving a single transition state have been characterised using
indices such as
\begin{equation}
\widetilde N = { \frac{ \left( \sum_t \left|{\bf X}_t(S)-{\bf X}_t(F)\right|^2 \right)^2}
{\sum_t \left|{\bf X}_t(S)-{\bf X}_t(F)\right|^4 }},
\end{equation}
which is a measure of the number of atoms that participate in the rearrangement.
Here ${\bf X}_t(S)$ and ${\bf X}_t(F)$ are the position vectors of atom $t$ in the
starting and finishing geometries, respectively.
The largest values are marked in Figure \ref{fig:tests} next to the corresponding
transition state.
It is noteworthy that the pathways LJ$_{38}$~$3.956\sigma$ and LJ$_{75}$~$4.071\sigma$
both involve some highly cooperative steps, and the average value of $\widetilde{N}$ is more than
12 for both of them.

We have found that it is usually easier to locate good transition state
candidates for a multi-step path
if the stationary points are separated by roughly equal distances, in
terms of the integrated path length.
Furthermore, it seems that more effort is needed to characterise a multi-step path
when transition states involving very different path lengths are present.
In such cases it is particularly beneficial to build up a complete path in stages.
To further characterise this effect we
introduce a path length asymmetry index $\pi$ defined as
$\pi = \vert s_+ - s_- \vert/(s_+ + s_-)$, where $s_+$ and $s_-$ are the two integrated path lengths
corresponding to the two downhill steepest-descent paths from a given transition state.
For example, in rearrangement LJ$_{38}$~$3.956\sigma$, five steps out of nine have $\pi > 0.5$.

Barrier asymmetry also plays a role in the accuracy of the tangent estimate, the
image density required to resolve particular regions of the path, and
in our selection process for transition state candidates, which is based on the condition $E_i > E_{i\pm1}$.
To characterise this property we define a barrier asymmetry index, $\beta$, as
$\beta = \vert E_+ - E_- \vert/max\left(E_+,E_-\right)$, where $E_+$ and $E_-$ are the barriers corresponding
to the forward and reverse reactions, respectively.
The test cases in Figure \ref{fig:tests} include a variety of situations,
with barrier asymmetry index $\beta$ ranging from $0.004$ to $1.000$. 
The maximum values of $\pi$ and $\beta$ are shown next to the corresponding transition states
in this Figure.

We note that the total number of gradient evaluations required to produce the above paths
could be reduced significantly by optimising the DNEB parameters or the connection strategy 
in each case. However, our objective was to find parameters that give reasonable results for
a range of test cases, without further intervention.

\section{Conclusions}
\label{sec:conclusions}

The most important result of this work is probably the doubly nudged elastic band (DNEB) formulation,
in which a portion of the spring gradient perpendicular to the path is retained.
With this modification we found that
L-BFGS minimisation of the images is stable, thus providing a significant improvement in efficiency.
Constraints such as elimination of overall rotation and translation are not required, and
the DNEB/L-BFGS method has proved to be reliable for relatively complicated cooperative
rearrangements in a number of clusters.

In comparing the performance of the L-BFGS and quenched velocity Verlet (QVV) methods for optimising 
chains of images we have also investigated a number of alternative QVV schemes.
We found that the best approach is to quench the velocity after the coordinate update, so
that the velocity response to the new gradient lags one step behind the coordinate updates.
However, this slow-response QVV (SQVV) method does not appear to be competitive with L-BFGS.

Finally, we have revised our previous scheme for constructing connections between distant 
minima using multiple transition state searches. Previously we have used an NEB/L-BFGS framework
for this purpose, with eigenvector-following refinement of transition state candidates and
characterisation of the connected minima using energy minimised approximations to the steepest-descent
paths \cite{wales02,evansw03}.
When the DNEB/L-BFGS approach is used we have found that it is better to spend more effort in the
DNEB phase of the calculation, since a number of good transition state guesses can often be
obtained even when the number of images is relatively small.
In favourable cases a complete path linking the required endpoints may be obtained in one cycle.
Of course, this was always the objective of the NEB approach
\cite{jonssonmj98,henkelmanj00,henkelmanuj00,henkelmanjj00},
but we have not been able to achieve such results reliably for complex paths without 
the current modifications.
When a number of transition states are involved we still find it more efficient to
build up the overall path in stages, choosing endpoints that become progressively closer in space.
This procedure has been entirely automated within the OPTIM program, which can routinely
locate complete paths for highly cooperative multi-step rearrangements, such as those 
connecting different morphologies of the LJ$_{38}$ and LJ$_{75}$ clusters.

\section{Acknowledgements}
S.A.T. is a Cambridge Commonwealth Trust/Cambridge Overseas Trust scholar. Most of the calculations
were performed using computational facilities funded by the Isaac Newton Trust.

\bibliography{001405jcp}
\clearpage
\section{Tables}

\begin{table}[h]
\caption{The minimal number of images and total number of gradient calls (in parentheses) are shown
for degenerate rearrangements of LJ$_7$.
The image range was $2 \le N \le 20$ and the iteration range was $1 < \ell \le 3,000N$.
Each SQVV calculation was started from the guess produced using linear interpolation, while
guesses for L-BFGS runs were preoptimised using DNEB/SQVV until the RMS force dropped below 2.0. Every iteration the
images that satisfy $E_i > E_{i\pm1}$ were optimised further using eigenvector-following. The
transition state candidates that converged 
to a true transition state within five iterations were used to generate the connected minima
using energy minimisation. If this procedure
yielded a connected pathway the calculation was terminated and the rest of the parameter range 
was not explored. Otherwise, the
number of images was incremented and the procedure repeated. The number
of gradient calls is a product of the number of images and the total
number of iterations. For the L-BFGS calculations the number of iterations includes the SQVV preoptimisation steps (100 on
average) and the actual number of L-BFGS steps. Dashes signify cases where we were unable to obtain a connected pathway.}
\label{tab:1}
\begin{center}
\begin{tabular}{lcccc}
\hline
\hline 
Method & 1--2 & 2--3 & 3--4 & 4--5 \\
\hline
DNEB/L-BFGS &  5(1720) & 18(30276)& 11(2486) & 18(8010) \\
DNEB/SQVV   & 16(21648) & --   & 10(14310)& --    \\
\hline
\hline

\end{tabular}
\end{center}
\end{table}

\begin{table}
\caption{The minimal number of iterations needed to produce connected pathways for four degenerate rearrangements of LJ$_7$
using a 50-image NEB. The strategy of this calculation is identical to the one described in the caption to Table \ref{tab:1},
except that the number of images was fixed.}
\label{tab:2}
\begin{center}
\begin{tabular}{lcccc}
\hline
\hline 
Method & 1--2 & 2--3 & 3--4 & 4--5 \\
\hline
DNEB/L-BFGS &   131$^a$ &   493 &   171 &   326    \\
DNEB/SQVV   &  1130 & 15178 &  2777 & 23405$^b$ \\
NEB/SQVV    & 11088$^b$ &     -- & 30627 &    --      \\
\hline
\hline
\end{tabular}
\end{center}

$^a$ The number of iterations is 
the sum of the SQVV preoptimisation steps (100 on average) and the actual number
of iterations needed by L-BFGS minimiser. 
$^b$ This value is not directly comparable since DNEB converged to a different
path that contains more intermediate minima. 
Dashes signify cases where we were unable to obtain a connected pathway.
\end{table}

\clearpage

\section{Figure Captions}
\begin{figure}
\centerline{\includegraphics{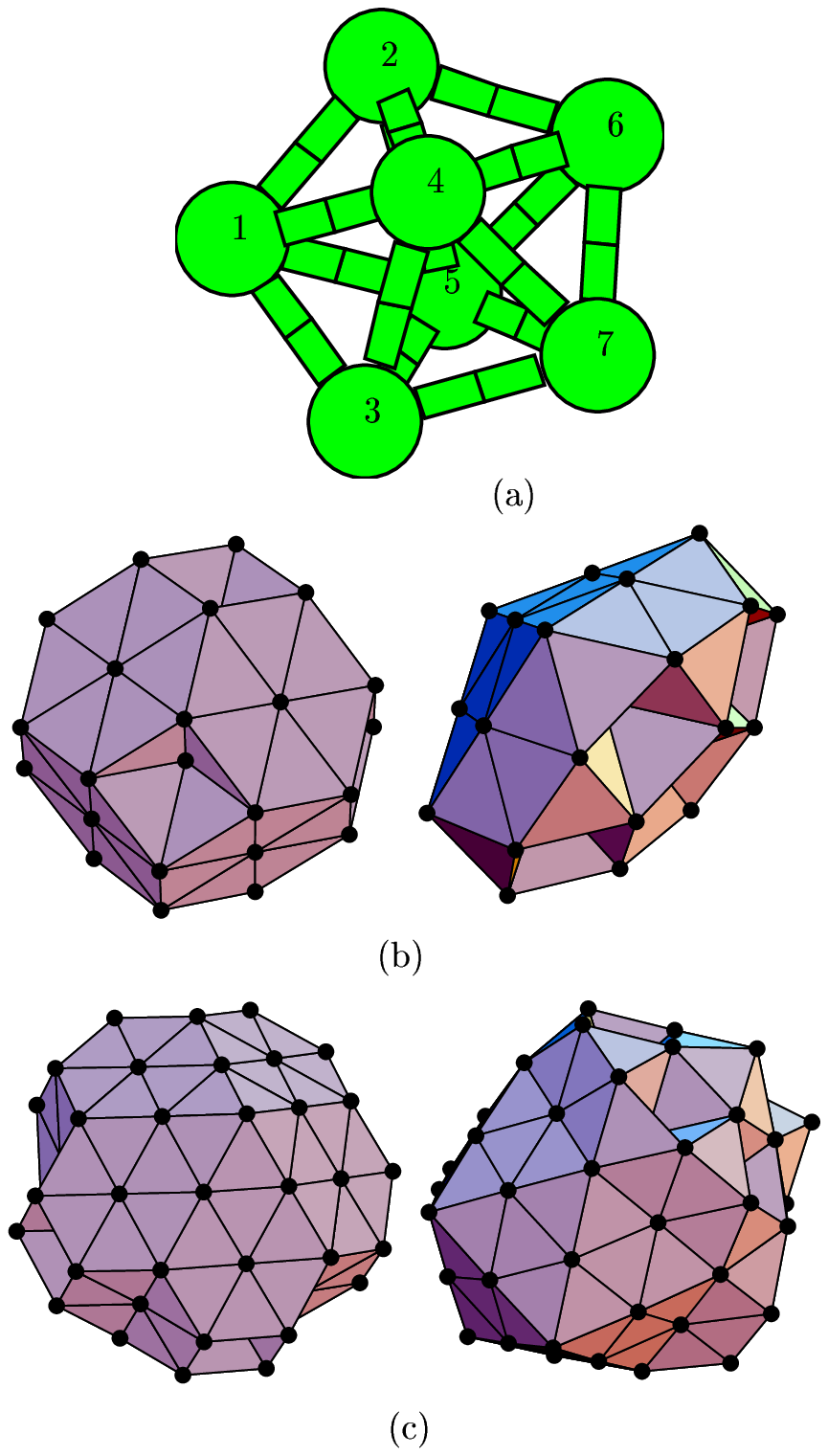}}
\caption{}
\label{fig:ep}
\end{figure}

\begin{enumerate}
\item
     Graphical representation of the nudged elastic band approach.
	(a) The optimised nudged elastic band for a two-dimensional model surface.
	The band contains $21$ images and connects two minima $\bfx_0$ and $\bfx_{23}$.
     Image $\bfx_9$ has the highest energy and might therefore be used to 
     estimate transition state properties or as a starting
     guess for further refinement.
	(b) `Nudging': the NEB depicted in (a) is projected onto the $xy$ plane and
     feels only the perpendicular component of the true gradient from the effective
	potential $V^{\perp}$.
\item
     Details of recent NEB implementations. (a) Conditions under which kinks appear
     during optimisation of the NEB using the tangent
     estimated from the line segment $\bft$ connecting images $i+1$ and $i-1$.
     Displacement of image $i$ from the path (dash-dotted line)
     creates forces ${\bf F}^{\perp}_{i-1} = - \gtper_{i-1}$ and ${\bf F}^{\perp}_{i} = - \gtper_i$.
     While ${\bf F}^{\perp}_{i}$ is a restoring force that originates from $V^{\perp}$,
     $ {\bf F}^{\perp}_{i-1}$ is destabilising and originates from
     $V^{\parallel}$ (and is non-zero due to the fact that the tangent at image $i-1$ has changed after displacement of
     image $i$).
     For the case of small displacements the potential may be resolved into two contributions,
	$V^{\perp}=(k^{\perp}/2)x^2$ and $V^{\parallel}=-k^{\parallel}y$, and
     kinks will not appear if $k^{\parallel}/k^{\perp} < \vert \bft \vert$.
     (b) Tangent estimate using the higher energy neighbour: image $i+1$
     is `hanging' on to image $i$.
     The separation $d$ is controlled by the lower-lying images $(> i+1)$ but not $V$. (c) An NEB
     that follows the curved region of the path: since the spring force ${\bf F}_1$ acting on image
	$i$ is compensated by projection ${\bf F}_2$, the distribution of images becomes uneven. (d) Corner-cutting
     displayed on a cross-section of the curved part of the path depicted in (c): the image is displaced from
	the path due to the presence of ${\bf F}_{spr}^{\perp}$.
\item
	(a) Number of iterations, $\ell$, and (b) average deviation from the average image separation, $\varsigma$,
	as a function of the spring force constant, $k_{spr}$, obtained using a 17-image NEB on the
     M{\"{u}}ller-Brown surface~\cite{mullerb79}. Minimisation was performed using QVV with time step $0.01$
	and RMS force termination criterion $0.01$.
     The number of iterations is shown for velocity quenching after the coordinate update (diamonds),
     after the gradient evaluation (squares) and at the half-step through the velocities update (stars). 
\item
	(a) Number of iterations, $\ell$, and (b) average deviation from average image
	separation, $\varsigma$, as a function of the spring force constant, $k_{spr}$, obtained
	using a 17-image NEB for the M{\"{u}}ller-Brown surface~\cite{mullerb79}. Minimisation was performed using
	L-BFGS with number of corrections $m=4$, maximum step size $0.1$
	and RMS force termination criterion $0.01$.
\item
	RMS gradient $g^{\perp}_{\text{RMS}}$ as a function of iteration number $\ell$. A 7-image NEB
	was used to model an isomerisation path in the LJ$_7$ cluster (global minimum $\rightarrow$
	second-lowest minimum). Minimisation was performed
	using the SQVV (a) and L-BFGS (b) methods. Results are shown for minimisations with and without
	removing overall rotation and translation (diamonds and stars, respectively).
	The inset in (a) depicts the average deviation from the average image
	separation, $\varsigma$, as a function of iteration number for minimisations using SQVV,
	while the inset in (b) shows $g^{\perp}_{\text{RMS}}$ recorded for $1000$
	iterations of L-BFGS minimisations. These calculations were all continued for a fixed number
	of iterations, regardless of convergence.
\item
	Structures of the most stable isomers for (a) LJ$_7$, (b) LJ$_{38}$ and (c) LJ$_{75}$ clusters,
        which were used as endpoints in the NEB calculations. 
        The first endpoint was the global minimum in each case.
        For LJ$_{38}$ and LJ$_{75}$ the second endpoint was chosen
	to be second-lowest minimum shown on the right in parts (b) and (c), respectively, while 
	a permutational isomer of the global minimum was used as the second endpoint in all the 
	LJ$_7$ calculations. 
        The notation 1--2 denotes an LJ$_7$ rearrangement
	where the second endpoint is structure (a) with atoms 1 and 2 swapped.
	The structures and numbering employed for LJ$_{38}$ and LJ$_{75}$
	are defined at http://www-wales.ch.cam.ac.uk/$\sim$sat39/DNEBtests/.
\item
	The energy, $E$, as a function of the integrated path length, $s$, for four
	degenerate rearrangements of LJ$_7$. These profiles were constructed
	using energy minimisation to characterise the paths connected to transition
	states obtained by EF refinement of candidate structures obtained
	from DNEB calculations~\cite{wales94a}.
\item
	The energy, $E$, as a function of integrated path length, $s$,
	for pathways linking the two lowest minima of LJ$_{38}$ and LJ$_{75}$.
	Calculations were initiated between two different sets of permutational
	isomers of these minima.	For each profile the number of transition states, $N_t$, number of DNEB
	runs, $N_c$, and the total number of gradient calls, $N_g$, are shown.
	Maximum values of $\widetilde{N}$, $\beta$ and $\pi$ are marked next to the
	corresponding transition states. The endpoints were illustrated in Figure \ref{fig:ep}.
\end{enumerate}
\clearpage

\clearpage

\clearpage

\clearpage

\clearpage

\clearpage

\clearpage

\clearpage

\clearpage

\end{document}